**Synthesis dependent characteristics of $Sr_{1-x}Mn_xTiO_3$ (x=0.03, 0.05, 0.07 and 0.09)**

K.R.S.Preethi Meher[1], C. Bogicevic[2], Pierre-Eymeric Janolin[2] and K.B.R.Varma[1*]

[1]*Materials Research Centre, Indian Institute of Science, Bangalore 560012*
[2]*Laboratoire Structures, Proprietes et Modelisation des Solides, Ecole Centrale Paris, CNRS-UMR8580, Grande Voie des Vignes, 92295 Chatenay-Malabry Cedex, France*

*Corresponding author, Email: kbrvarma@mrc.iisc.ernet.in ;
Fax: 91-80-23600683 Tel No: 91-80-22932914

**Abstract**

$Sr_{1-x}Mn_xTiO_3$ (where x = 0.03, 0.05, 0.07 and 0.09) was synthesized via different routes that include solid-state, oxalate precipitation and freeze drying. In oxalate precipitation technique, compositions corresponding to 3 and 5 mol % doping of Mn were monophasic whereas the higher compositions revealed the presence of the secondary phases such as MnO, $Mn_3O_4$ etc., as confirmed by high resolution X-ray diffraction (XRD) studies. The decomposition behavior of the precursors prepared using oxalate precipitation method corresponding to the above mentioned compositions was studied. Nanopowders of compositions pertaining to 5 to 9 mol % of Mn doping were obtained using freeze-drying technique. The average crystallite size of these nanopowders was found to be in the 35 to 65 nm range. The microstructural studies carried out on the sintered ceramics, fabricated using powders synthesized by different routes established the fine grained nature (< 1 µm) of the one obtained by freeze drying method. Raman scattering studies were carried out in order to complement the observations made from XRD regarding the phase purity. The dielectric properties of the ceramics obtained by different synthesis routes were studied in the 80 – 300 K temperature range at 100 kHz and the effect of grain size has been discussed.

**Keywords:** $SrTiO_3$, chemical synthesis, Raman spectroscopy, dielectric constant

## 1. Introduction

SrTiO$_3$ (STO) ceramics were known for their promising applications in the design and fabrication of multilayer capacitors and varistors [1] while the thin films of STO have been studied extensively for their possible use in microwave filters [2]. Recently there has been renewed interest concerning the study of the physical properties of STO with different cation doping (Mn, Fe, Cr, Co etc) in view of its importance for magnetoelectric effect based applications. Shravatsman et al [3] reported that Sr$_{0.98}$Mn$_{0.02}$TiO$_3$ behaves like a dipolar glass and a spin glass with their corresponding transitions lying around 38 K and 34 K respectively, which has drawn the attention of many researchers. At this juncture, it is worth mentioning that Smolenskii [4] encountered with a dielectric anomaly in the 20-30 K temperature range in STO. However this observation was in complete contrast with that of of Hulm et al [5] where no such anomaly was observed for STO down to 1.3 K with the dielectric constant values saturating at ~ 1300 below 4 K. This discrepancy was later addressed to by Smolenskii [6] and attributed it to the presence of impurities and the method via which the samples are synthesized. Later, low temperature dielectric measurements carried out on STO single crystals showing a rise in the dielectric constant values from 370 (at RT) to ~18000 (1.4 K) was reported by Weaver [7]. Also, ferroelectric hysteresis loops were recorded for STO single crystals below 35 K indicating the switching of spontaneous polarization though the observed values were as low as 1.5 µC/cm$^2$ at 1.4 K [7]. In order to unravel this mystery, Smolenskii [8] carried out a systematic investigation into the influence of various levels of Bi-doping on the dielectric behavior of STO. This was followed by many similar studies [9, 10] which revealed the role of Bi substitution in inducing strong and multiple relaxation modes. Notable results were also obtained for Ca doped STO where low temperature ferroelectric-like phase transitions could be induced by random field interactions between Ca off-centered polar clusters [11]. The effect of dopants and various oxygen partial pressures (maintained during the synthesis) in altering the onset of phase transition temperatures in STO along with its overall dielectric behavior have become a topic of current interest [12-17].

Indeed interesting magnetic characteristics such as spin glass behavior that was reported for $Sr_{0.98}Mn_{0.02}TiO_3$ has provided enough motivation for a systematic study on magnetic ordering with various levels of Mn doping in STO [18, 19]. Keeping the observations and contradictions in view, we felt that it was important to synthesize phase pure STO doped with Mn. A survey of the earlier work demonstrates that $SrTiO_3$ could be prepared by various physical and chemical routes that include conventional solid-state reaction route, chemical-coprecipitation method, sol-gel, hydrothermal and electrochemical [20-25]. Bera et al [26] reported a simple technique for the formation of $SrTiO_3$ where the strontium oxalate is precipitated over the fine $TiO_2$ particles (which facilitate seeding for heterogenous nucleation) followed by subsequent heat treatment at appropriate temperatures. Apart from these methods, oxides which are prepared by freeze drying has a reduced probability of the formation of secondary phase usually resulting from the inhomogeneity of the starting powders and also have better sinterability compared to that of the other methods. Many known complex oxides have been prepared using this method in which the resultant powders could be obtained in an amorphous state [27]. For the present investigation, STO powders doped with different amounts of Mn were synthesized by three different routes. We illustrate the details of the synthesis and characterization of $Sr_{1-x}Mn_xTiO_3$ (x= 3, 5, 7 and 9 mol % abbreviated as SMTO3,

SMTO5 SMTO7 and SMTO9 respectively) powders obtained via solid-state reaction route, chemical co-precipitation of oxalates and freeze drying methods.

## 2. Experimental procedure

### 2.1 Solid-state reaction method

The polycrystalline powders of the composition SMTO3 were prepared by conventional solid-state synthesis route as the current interest is to visualize the influence of higher contents of Mn doping on the physical properties of SMTO. For this, $SrCO_3$, MnO and $TiO_2$ were mixed homogeneously by subjecting them to ball milling for 12 hours. The ball milled powders were heated at 1073 K for several hours in order to ensure the decomposition of carbonates and subsequently calcined at 1273 K for 5 h. Later the calcined powders were pressed into pellets and sintered at 1623 K for 24 h ambient atmospheres.

### 2.2 Co-precipitation of oxalates

The starting precursors used for the synthesis were Strontium Chloride hexahydrate $(SrCl_2).6H_2O$ (>99% assay from Aldrich), Manganous chloride tetrahydrate $(MnCl_2.4H_2O)$ (> 99% assay from Merck) and $TiO_2$ (<1 µm and >99% assay from Merck). An aqueous solution was obtained by dissolving stoichiometric amounts of $SrCl_2.6H_2O$ and $MnCl_2.4H_2O$ in deionized water. Subsequently the required $TiO_2$ powder (such that Sr/Ti=1.0) was added to the aqueous solution under constant stirring. Vigorous stirring and ultrasonication of the solution was carried out to break the bigger $TiO_2$ agglomerates into fine particles in the solution. The required amount of oxalic acid $(C_2H_2O_4$; Analytical Grade) to obtain 0.1 M standard solution was dissolved in water and the solution was added drop by drop to the aqueous solution. This led to the precipitation of Strontium and Manganous oxalate onto the fine $TiO_2$ particles suspended in the solution. The precipitate was washed with deionized water for several times until all the chlorides were removed and was recovered and dried in an electric oven maintained at 423 K. The dried powders were collected from the beaker for further characterization. The powders were subsequently calcined in air in the 1073 to 1373 K temperature range with intermittent grinding to ensure homogenization. XRD studies were carried out on the calcined powders in order to assess the desired phase formation. The procedure for synthesizing SMTO powders using oxalate precipitate method is illustrated in the form of a block diagram in Fig. 1.

## 2.3. Freeze-drying method

The principle of freeze-drying method is based on the removal of water by sublimation of the frozen solution containing the cations in the stoichiometric ratio. The factors that have to be taken care of during freeze drying are (a) chemical compatibility between the salts used in order to avoid precipitation after their dissolution and (b) determination of solution freezing point, which depends on the type of solvent (such as water, alcohol etc) used for synthesis other than the molar concentration of the solute [27]. In order to synthesize SMTO, Strontium acetate ($Sr(C_2H_3O_2)_2.0.5H_2O$, > 99.5 % from Aldrich), Titanium isopropoxide (Ti{OCH(C_4 )} >99.5 % from Aldrich) and Manganese acetate ($Mn(CH_3COO)_2 \cdot 4 H_2O$, >99.5% from Aldrich) were used. Stoichiometric amounts of Strontium acetate and Manganese acetate were dissolved together in 200 ml of deionized water. Titanium isopropoxide was poured into a beaker containing a mixture of isopropyl alcohol ($C_3H_8O$, 99.5 % from Prolabo) and acetic acid (99.8% from Prolabo) in 1:3 volumetric proportions to avoid hydrolysis. This solution was then added to that containing the other two dissolved salts. The resulting solution was mixed thoroughly using a magnetic stirrer and maintained at a temperature of 373 to 423 K. The clear boiling solution is then sprayed into a liquid nitrogen bath. The frozen mass was recovered from the bath and dried on a plate kept inside a freeze-drier (ALPHA 2-4, Christ) under vaccum (3.5 Pa). Initially the temperature of the powder was maintained at 253 K under a low vaccum for 24 h where the water is removed by sublimation and then the temperature was gradually increased upto 343 K till the powders became devoid of moisture. The fluffy recovered powders were then calcined in the 973 K to 1173 K temperature range and the phase formation was investigated by XRD studies. The procedure adopted to synthesize SMTO by freeze drying method has been presented in the form of a block diagram in Fig. 2.

The powders that were obtained by both the above synthesis methods were isostatically cold-pressed upto 750 MPa and the compacted pellets were sintered in the temperature range of 1473 to 1623 K. The density of the pellets, measured by employing Archimedes method, was in the range of 91-95 %.

## 2.4 Characterization methods

The decomposition behavior of the precursor powder obtained from oxalate precipitation method was studied by subjecting it to differential thermal analysis (DTA) and Thermo gravitometric analysis (TGA, Thermofisher Scientific) at a heating rate of $10^o$C/min in air in the 323 to 1473 K temperature range. X-ray diffraction studies were carried out using a high resolution diffractometer set in Bragg-Breneto geometry and operated with a rotating anode generator (Rigaku RU300 operated at 18 kW) using Cu K$\alpha$ radiation along with Siemens detector. Scanning Electron Microscopy associated with backscattering mode and Energy dispersive X-ray analysis (EDAX) (SEM, Quanta ESEM) of the above samples have been done in order to shed more light on their microstructure and local chemical composition. Raman measurements were carried out using Micro-Raman spectrometer (LanRAM HR800, Horiba Jobin Yvon) with an Ar ion laser excitation source (excitation wavelength of 514.5 nm) and the illuminated spot size was kept at 0.5-0.7 µm. Dielectric measurements were carried out using Agilent 4210 high precision impedence analyzer, on the cylindrical sintered pellets sputtered with gold electrodes down to 80 K in the 100 Hz – 1 MHz frequency range at an applied voltage of 0.5 V ($V_{r.m.s}$).

## 3. Results and Discussion

### 3.1 Structural, Microstructural and Thermal decomposition studies

The high resolution X-ray diffraction (HXRD) pattern of the powders derived from the bulk ceramics of SMTO3 (sintered at 1623 K for 12 h) obtained by solid-state reaction route is shown in Fig. 3. The peaks were indexed based on the space group P*m*-3*m* confirming the formation of the cubic perovskite phase. However the extra peaks present (inset of Fig. 3) along with the peaks corresponding to the major phase could be indexed to $Mn_3O_4$. The amount of $Mn_3O_4$ was quantified to be roughly upto 0.5 % based on the XRD peak intensity ratio. In order to bring down the sintering temperatures which were thought to be linked up with phase purity, we directed our efforts to synthesize the present compound with varied Mn contents (3 to 9 mol %) by oxalate precipitation technique.

DTA and TGA traces obtained for the precursor powders of SMTO in the oxalate precipitate method are depicted in Fig. 4. As this synthesis method involves the co-precipitation of the Strontium oxalate and Manganese oxalate on to the $TiO_2$ powders, the distinct thermal decomposition steps corresponding to each of these precursors could be observed in TGA. The small weight loss initially observed at less than 373 K and the corresponding endothermic peak in DTA curve is due to the removal of moisture from the sample. The steps I and II involve a total weight loss of $\Delta m \sim 7\%$ and the two corresponding endothermic peaks occurring in the temperature range of 373 – 473 K are due to the removal of water molecules from Strontium oxalate ($SrC_2O_4 \cdot H_2O$) and Manganese oxalate ($MnC_2O_4 \cdot 2H_2O$). As the decomposition peaks of Manganese oxalate overlap with that of Strontium oxalate, only the total weight loss could be considered during the analysis. The next weight loss step (step III) corresponding to the exothermic peak in DTA is due to the decomposition of Strontium oxalate and Manganese oxalate to their respective carbonates with the evolution of CO molecules [28]. The final weight loss step (step IV) corresponding to the weak endothermic peak is due to the liberation of $CO_2$ and subsequently the residues i.e., respective oxides left back react with $TiO_2$ resulting in the formation of the desired phase. The general equations representing the above reactions and the respective $\Delta m$ are given below.

Step I & II:
$$SrC_2O_4 \cdot H_2O \xrightarrow{373-473K/endo} SrC_2O_4 + H_2O$$
$$MnC_2O_4 \cdot 2H_2O \xrightarrow{373-473K/endo} MnC_2O_4 + 2H_2O$$

Total $\Delta m \sim 7.2 - 8.3\%$

Step III:
$$SrC_2O_4 \xrightarrow{573-843K} SrCO_3 + CO$$
$$MnC_2O_4 \xrightarrow{573-843K} MnCO_3 + CO$$
$$CO + 1/2 O_2 \xrightarrow{O_2/exo} CO_2 \uparrow$$

Total $\Delta m \sim 7.2 - 8\%$

Step IV: $(1-x)SrCO_3 + xMnCO_3 + TiO_2 \xrightarrow{1123-1353K/endo} Sr_{1-x}Mn_xTiO_3 + CO_2 \uparrow$

Total Δm~10 – 11.2 %

The formation of phases, synthesized using the oxalate precipitate technique, was confirmed later by high resolution X-ray diffraction technique. XRD powder patterns of the sintered samples of SMTO are depicted in Fig. 5 (a-f). The XRD patterns of the first two compositions such as SMTO3 and SMTO5 confirmed the formation of a pure perovskite phase associated with the space group P*m*-3*m*. However, symptoms of the presence of the secondary phases such as MnO and $Mn_3O_4$ were found for x > 5 mol. The X-ray peak positions reported in the literature [29, 30] for MnO and $Mn_3O_4$ have been given for the purpose of comparison in Fig. 5(a) and 5(b) respectively. Fig.5(c-f) shows the high resolution XRD powder patterns obtained for sintered samples of SMTO3, SMTO5, SMTO7 and SMTO9 respectively. The high intensity peak of MnO at 2θ=40.54° is close to that of the extra peak (though not very evident in Fig. 5(e)) occurring in the XRD pattern of SMTO7 case confirming its presence as a secondary phase (Fig. 5 (d)). In the case of SMTO9 along with MnO and $Mn_3O_4$, $TiO_2$ peaks (JCPDS No.118875) were also found to be present. In order to have better illustration of the phase purity of SMTO3 and SMTO5 prepared using the oxalate precipitate method, their indexed XRD patterns have been plotted in logscale as shown in Fig 6 (a-b). In order to enhance the solid solubility limit and to achieve mono phasic compounds devoid of impurities, it is essential to control the homogeneity of the starting powders. One of the alternate synthesis methods to obtain homogeneous compositions was thought to be freeze drying. We have adopted this method to synthesize STO doped with various levels of Mn.

The amorphous nature of the as-synthesized powders obtained from freeze drying method is directly evident from recorded XRD pattern given in the inset of Fig. 7. The as synthesized powders have been subjected to calcination and the XRD patterns recorded for the calcined powders of SMTO5 ( 1073 K for 3 h), SMTO7  (1073 K for 3 h  and SMTO9 (1103 K for 3 h) compositions are shown in Fig. 7(a-c)

The crystallite size was calculated using the well known Scherrer formula $D = \frac{0.9\lambda}{\beta \cos\theta}$ , where D is the particle size, λ is the wavelength of the X-rays, β is the full width at half maximum which is corrected for the instrumental error using a standard sample of $LaB_6$ and θ is the angle at which the diffraction peak occurs. The crystallite sizes calculated using XRD data of calcined powders of STMO are listed in Table 1.

The calcined powders were sintered in order to obtain bulk ceramics suitable for further characterization. The sintering was carried out in the temperature range of 1073 K to 1546 K. Detailed XRD studies of the STMO5, STMO7 and STMO9 ceramics sintered at different temperatures have been carried out and the results obtained are shown in Fig 8(a-c) respectively. Though traces of secondary phase corresponding to $Mn_3O_4$ was found to be present in both SMTO5 and SMTO7, the sintering temperatures could be optimized such that the bulk samples were devoid of any secondary phases as shown in the Figs.8(a) and (b). In the case of SMTO5, sample sintered at 1473 K was found to have $Mn_3O_4$ as an impurity phase. Interestingly, on increasing the sintering temperature up to 1523 K there is a decrease in the content of $Mn_3O_4$. While in the case of SMTO7, the samples sintered at 1473 K was found to be monophasic with no detectable traces of $Mn_3O_4$ and with increasing sintering temperature, the traces of $Mn_3O_4$ started appearing. In the case of SMTO9, significant amount of secondary phases such as $Mn_3O_4$ and MnO were found to exist in the bulk samples in spite of annealing over a range of high temperatures. We did not find significant changes in the formation of the impurity phases such as MnO and $Mn_3O_4$ with a change in annealing time. Tkach et al [31] reported the XRD patterns of $Sr_{1-x}Mn_xTiO_3$ where x=0.02, 0.03, 0.05, 0.1, 0.15 and 1 prepared by solid-state reaction route and reported the formation of $MnTiO_3$ impurity phase for the compositions $x \geq 0.1$. But in the present work, irrespective of the synthesis routes that were adopted, we always encountered with $Mn_3O_4$ as the impurity phase although MnO could also be found in the case of SMTO9. In fact the Electron Paramagnetic Resonance (EPR) studies carried out on $Sr_{1-x}Mn_xTiO_3$, where x is up to 3 mol %, by Badalyan et al [32] showed that the formation of Manganese oxide nanoparticles was favored before the incorporation of Mn into the STO lattice. In fact, the broader X-ray peaks corresponding to $Mn_3O_4$ corroborate their EPR studies.

Microstructural aspects of the sintered samples fabricated using both the oxalate precipitate and freeze drying methods have been studied in detail. Fig. 9 (a to c) shows the microstructures of the sintered samples of SMTO3, SMTO5 and SMTO7 prepared using the oxalate precipitate method. The average grain size associated with these samples is in the range of 1-2 μm. In the case of SMTO9, the grains of different morphologies were found and a spot EDAX on them clearly indicate phase segregation of a Ti rich secondary phase consistent with XRD data (inset of Fig.9 (d)). This is evident from the quantitative analysis of the elements present in the segregated and the major phase in SMTO9 as presented in Table 2. The homogeneity of the sample was also verified by carrying out EDAX analysis on the different regions of the sample. The quantitative estimation of the elements in regions 1 and 2 has also been listed in Table 2. The difference in the compositions of Mn in both the regions 1 and 2 is evident of the composition fluctuations present in these systems. The microstructures of the pellets fabricated using the powders obtained by freeze drying method was investigated using SEM. Fig. 10 (a and b) shows the high magnification images of SMTO5 and SMTO7 respectively. The observation of grains < 1 µm in size despite high synthesis temperatures involved points towards obtaining fine grained ceramics in the case of freeze-drying method as compared to that of the other methods. Fig. 10 (c) shows the microstructure of SMTO9 and the contrast in the corresponding back scattering image (given in Fig 10 (d)) clearly shows the presence of a secondary phase as envisaged in the XRD data.

### 3.2 Micro-Raman spectroscopic studies

Micro Raman studies carried out by earlier workers on the 0.1 at % of Mn doped STO at room temperature facilitated the identification of the presence of secondary phases such as $Mn_3O_4$ and $MnO_2$ [14]. Pure STO has a cubic phase at room temperature (space group P*m*-3*m*) and therefore there could be no first-order Raman active modes except the signature of second order scattering. However Rabuffetti et al reported first order modes in $SrTiO_3$ nanocubes that arise from loss of crystal structure symmetry caused by surface defects or impurities [33]. Therefore it might be interesting to investigate into the local structure of SMTO by using a powerful tool like Raman scattering. Fig. 11 shows the micro Raman spectra recorded for SMTO3, SMTO5, SMTO7 and SMTO9 compositions prepared by oxalate method. The overall spectra have mostly features from second order scattering similar to that of STO. The multiple phonon modes that give to rise to the second order scattering in STO around room temperature have been clearly identified and listed in the literature [34-36]. In the present work, the Raman spectra obtained for different compositions of STMO (synthesized by oxalate method) were corrected for the background and the peaks in the range of 100 cm$^{-1}$ to 900 cm$^{-1}$ was fit with Lorentzian function. The exact peak positions (the ten different modes identified are marked with alphabets from A-I in Fig. 11) were obtained and tabulated in Table 3 for all the compositions along with that of pure STO [34] for comparison. Further, Raman spectra obtained for SMTO5, SMTO7 and SMTO9 compositions (as shown in Fig. 12) synthesized using freeze drying method were also similarly analyzed and the peak positions have been listed in Table 3.

Raman spectra corresponding to SMTO3 and SMTO5, prepared by oxalate precipitation method, did not indicate any additional peaks arising due to impurities confirming our earlier observation of its phase purity based on X-ray diffraction studies. However the asymmetric character of the modes and their deconvolution (specifically for the modes C and D as reported in Table 3) into multiple peaks are interesting to ponder. The splitting of the first order scattering modes further with the appearance of sharp second order modes due to the breaking of crystal structure symmetry could be observed only at lower temperatures for STO [34]. However in the case of SMTO, weak splitting of modes could be observed even around room temperature that might indicate the local structural distortion induced by Mn substitution. The Raman spectra belonging to SMTO9 sample as shown in Fig. 11 has signatures of the impurity phase in the form of weak lattice mode of TiO$_2$ (strong intensity E$_g$ mode at 143 cm$^{-1}$ for pure TiO$_2$ [37]) observed at 141 cm$^{-1}$. The other dominating lattice mode observed at 433 cm$^{-1}$ could not be assigned to any known entity. In the case of ceramics fabricated by freeze drying method, SMTO9 has strong signature of the presence of $Mn_3O_4$ [38] (marked in Fig. 12(c)) as revealed by a weak second order mode observed at 661cm$^{-1}$. Similar observation of the asymmetric multiphonon scattering modes which might indicate the local structural distortion holds good for the samples prepared by freeze drying method also. In order to have further insight into this, one needs to carry out temperature dependent Raman scattering studies as in the case of pure STO.

**3.3 The effect of grain size on the dielectric behavior of SMTO**

The dielectric relaxation of the bulk ceramics of SMTO, synthesized by oxalate precipitation technique and the freeze drying method, have been studied in the (100 Hz – 1MHz) frequency range from 300 K down to 80 K. Variation of dielectric relaxation characteristics with various levels of Mn doping have been discussed elsewhere in detail [39]. In the present work, we focus on the dielectric characteristics of SMTO ceramics obtained by different syntheses routes. As no dispersion was observed in the dielectric constant values recorded in the 100 Hz – 1 MHz frequency range, the dielectric constant ($\varepsilon_r'$) and the loss (D) values obtained for 100 kHz have been listed for all the compositions fabricated by solid-state reaction, oxalate precipitation and the freeze drying methods (Table 4). The dielectric constant ($\varepsilon_r'$) values of the ceramics obtained by freeze drying method were observed to be almost half of that observed for the ceramics obtained by the solid-state reaction and the oxalate precipitation routes. These results evidently demonstrate the microstructure dependent dielectric property. It is to be noticed that the ceramics obtained by freeze drying consists of fine grains (with grain sizes lower than 1 µm) as compared to that (around 1-2 µm) obtained by the other routes.

This is further illustrated by the direct comparison of the temperature dependent dielectric constant behavior for different compositions prepared by oxalate precipitation (CG – coarse grained) and freeze drying (FG - fine grained) as given in Fig.13 . In fact, a similar trend has been observed consistently in undoped STO by Petzelt et al [40] when dielectric characteristics were studied for ceramics consisting of different grain sizes ranging from submicron to a few micrometers. The significant decrease in the observed dielectric constant values with decrease in the grain size was attributed to the dead layer effect. The thin grain boundary layer of very low dielectric constant acts as a dead layer and when its volume becomes comparable to that of the grains, the effective dielectric constant drastically decreases. In addition, it might be worth recalling that the bulk grain composition exhibits certain degree of fluctuation due to the inhomogeneous distribution of Mn (Table 2). This factor directly influences the pattern in the local distortion induced by the substitution of Mn in the Ti-site. Therefore the contribution of bulk composition fluctuations to the variation in dielectric behavior of SMTO ceramics synthesized by both oxalate precipitation and freeze-drying methods cannot be ruled out.

## 4. Conclusions

SMTO phase with 3 mol % Mn doping on the Strontium site could not be obtained phase pure when the solid state synthesis method was adopted. The oxalate precipitate method resulted in monophasic SMTO3 and SMTO5 while it was not useful for obtaining higher compositions of Mn (> 5 mol %). The freeze drying method was successfully used to synthesize high Mn containing compositions such as 5, 7, 9 mol % wherein amorphous starting powders were synthesized and heat treated to obtain bulk samples. Though, it was possible to achieve fine grained ceramics using the powders synthesized by freeze drying technique, the dielectric constants are lower than those obtained by the other methods. The temperature dependent dielectric behavior (from 80 K to 300 K) of the coarse grained and the fine grained ceramics fabricated from the different synthesis methods, was compared. A comparison of the multiphonon Raman scattering modes observed in the case of $Sr_{1-x}Mn_xTiO_3$ with that of pure $SrTiO_3$ seemed to uphold the possibility of local structural distortion.

## 5. Acknowledgement

One of the authors K.R.S.P would like to thank the European commission – External cooperation window for offering Erasmus Mundus fellowship and CNRS-SPMS, Ecole Centrale Paris, France for hosting her as visiting student.

**Table 1** The calculated particle size of the SMTO powders synthesized by freeze-drying method

| Compositions | crystallite size (nm) |
|---|---|
| SMTO5 | 65.2±0.1 |
| SMTO7 | 42.7±0.1 |
| SMTO9 | 35.9±0.1 |

**Table 2** Quantification analysis of various elements present in SMTO9 ceramics fabricated using oxalate precipitation method

| Segregated phase (secondary phase) | Element | Wt (%) | At (%) |
|---|---|---|---|
| | O (K) | 36.1 | 64.13 |
| | Sr (L) | 7.29 | 2.36 |
| | Ti (K) | 55.51 | 32.93 |
| | Mn (K) | 1.1 | 0.57 |
| Major phase (region 1) | O (K) | 20.85 | 53.15 |
| | Sr (L) | 53.11 | 24.73 |
| | Ti (K) | 25.43 | 21.66 |
| | Mn (K) | 0.61 | 0.46 |
| Major phase (region 2) | O (K) | 13.56 | 40.23 |
| | Sr (L) | 57.36 | 31.08 |
| | Ti (K) | 28.05 | 27.8 |
| | Mn (K) | 1.03 | 0.89 |

**Table 3** List of scattering modes (marked in Figs. 11 and 12) and the corresponding wavenumber values obtained from the multipeak fit for SMTO

| Multiphonon scattering modes as marked in Fig. 11 and 12 | Observed bands in SrTiO$_3$ (from Ref. (22) | Frequency shift (cm$^{-1}$) for SMTO fabricated by oxalate precipitation | | | | Frequency shift (cm$^{-1}$) for SMTO fabricated by freeze drying method | | |
|---|---|---|---|---|---|---|---|---|
| | | SMTO3 | SMTO5 | SMTO7 | SMTO9 | SMTO5 | SMTO7 | SMTO9 |
| A | 83 | 85 | 84 | 85 | 84 | 86 | 83 | 85 |
| B | 170 | 172 | 177 | 188 | 203 | 177 | 184 | 179 |
| C | 252 | 237<br>251<br>288 | 236<br>251<br>280 | 239<br>246<br>290 | -<br>244<br>283 | 235<br>249<br>278 | 235<br>250<br>286 | 236<br>251<br>294 |
| D | 316 | 324 | 310<br>347 | 341 | 321 | 300<br>321 | 315 | 323 |
| E | 380 | 374 | 384 | 384 | 386 | 365 | 374 | 366 |
| F | 620 | 619 | 617 | 620 | 612 | 610 | 625 | 616 |
| G | 678 | 671 | 683 | 686 | 688 | 695 | 683 | 687 |
| H | 725 | 722<br>751 | 727<br>749 | 715<br>752 | -<br>747 | 731<br>753 | 726<br>755 | 723<br>751 |
| I | | 795 | 793 | 795 | 806 | 796 | 794 | 794 |

**Table** 4 List of dielectric constant values ($\varepsilon_r'$) and loss (D) at 100 kHz of SMTO bulk ceramics fabricated by different synthesis routes

| Composition | $\varepsilon_r'$ (300 K) | D (300 K) | $\varepsilon_r'$ (80 K) | D (80 K) |
|---|---|---|---|---|
| solid-state method | | | | |
| SMTO3 | 261 | 0.0015 | 1582 | 0.011 |
| Oxalate precipitation method: | | | | |
| SMTO3 | 220 | 0.005 | 1110 | 0.01 |
| SMTO5 | 231 | 0.009 | 1069 | 0.008 |
| SMTO7 | 248 | 0.002 | 1371 | 0.006 |
| Freeze-drying method | | | | |
| SMTO5 | 90 | 0.0014 | 240 | 0.002 |
| SMTO7 | 122 | 0.002 | 418 | 0.0021 |
| SMTO9 | 107 | 0.002 | 308 | 0.0025 |

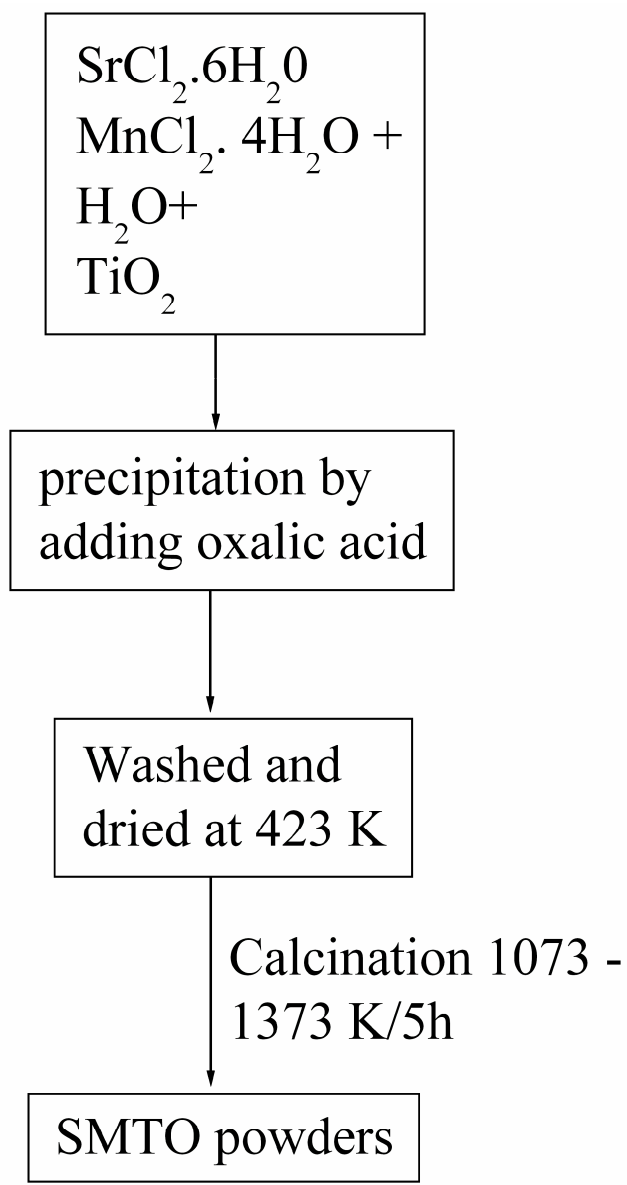

**Fig.1** Block diagram illustrating the steps followed during the oxalate precipitation synthesis of SMTO

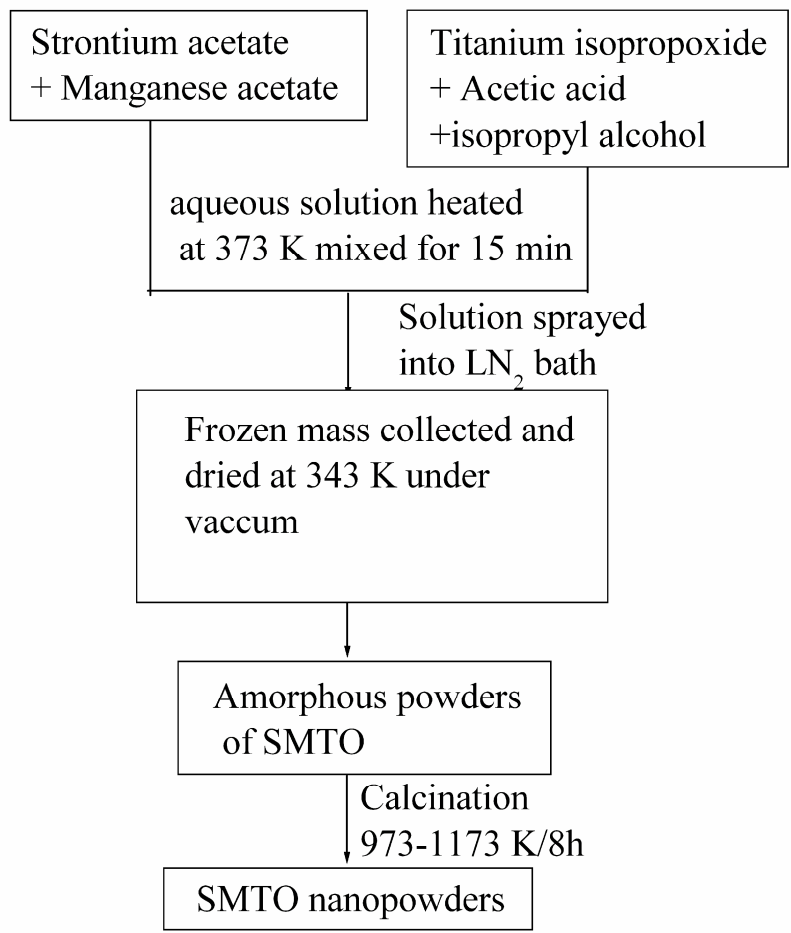

**Fig.2** Block diagram illustrating the methods followed during the freeze-drying synthesis of SMTO

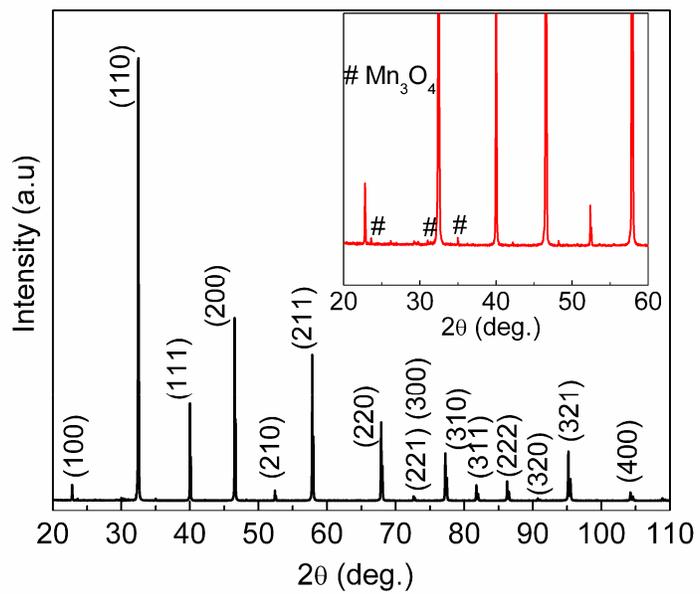

**Fig. 3** HXRD pattern for SMTO powders of the bulk ceramics sintered at 1623 K/ 12 h

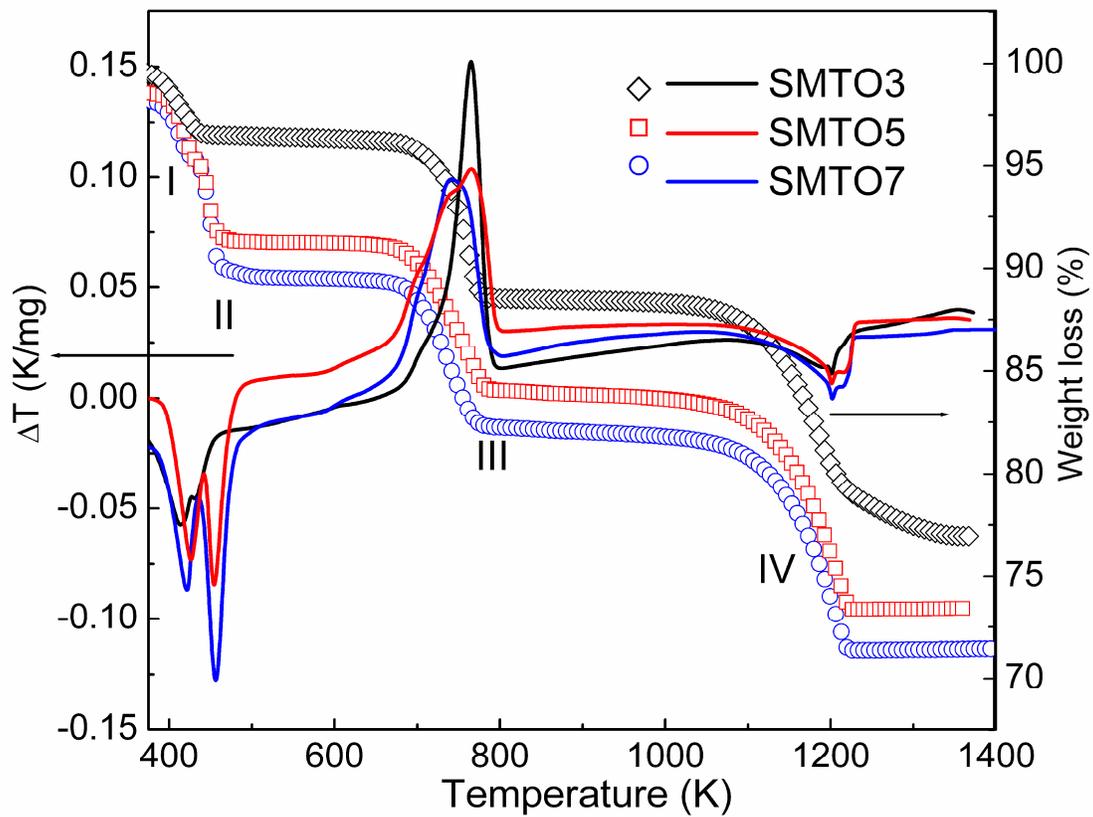

**Fig. 4** DTA and TGA curves of the as-synthesized starting powders corresponding to the compositions (a) SMTO3 (b) SMTO5 and (c) STMO7

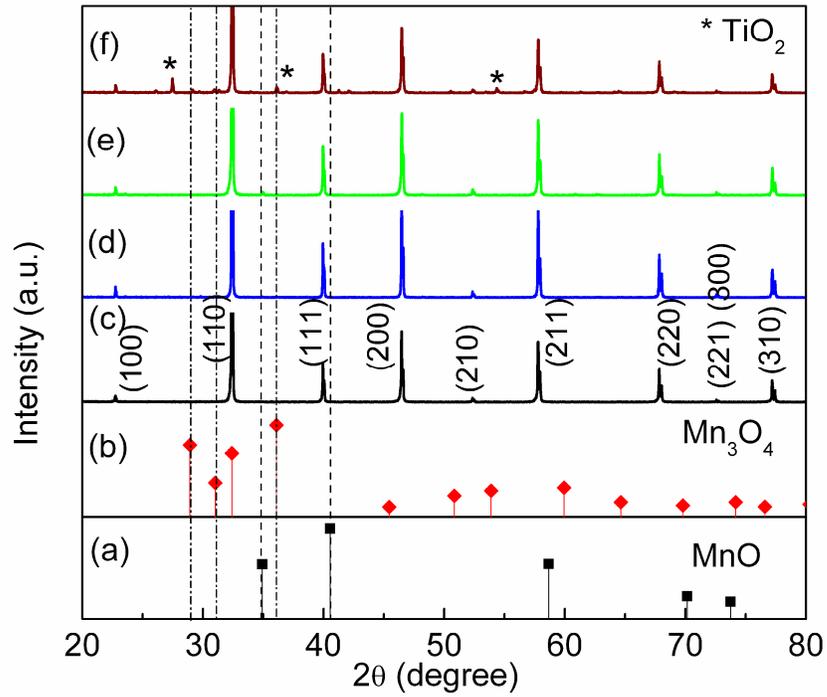

**Fig. 5** Peak positions of (a) MnO (filled squares) and (b) $Mn_3O_4$ (filled diamond) from standard data and XRD patterns of (c) SMTO3 (d) SMTO5 (e) SMTO7 (f) SMTO9 prepared from oxalate precipitation method. The dashed lines and the dash dot lines indicate the matching of peak positions of MnO and $Mn_3O_4$ respectively to the XRD data in (c) to (f)

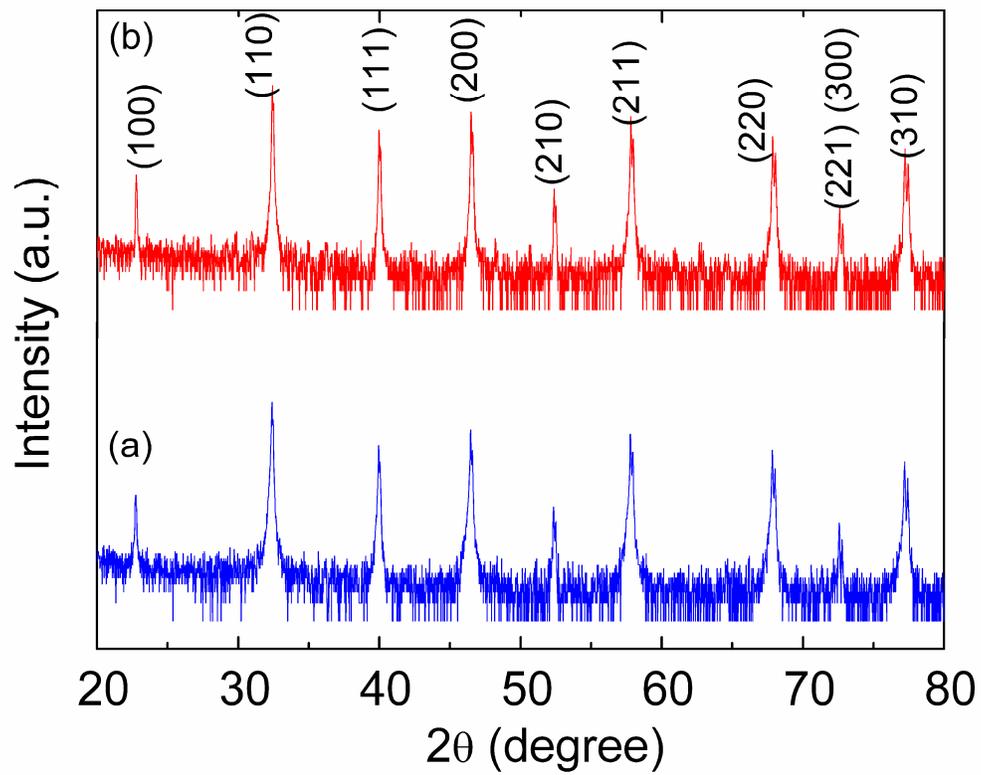

**Fig 6** High resolution XRD patterns of **(a)** SMTO3 and **(b)** SMTO5 plotted in logscale

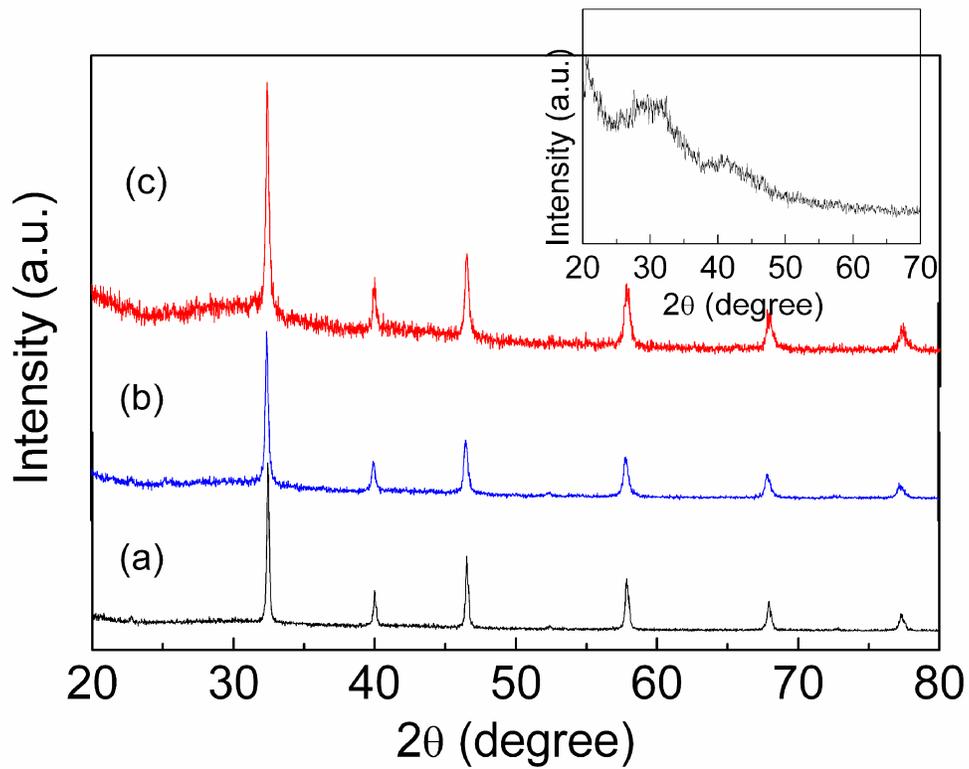

**Fig 7** XRD patterns of **(a)** SMTO5 **(b)** SMTO7 and **(c)** SMTO9 obtained after calcination and the inset shows the as synthesized powders obtained immediately after freezedrying procedure.

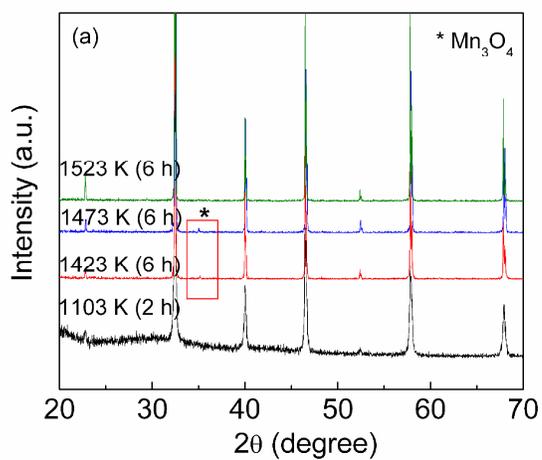

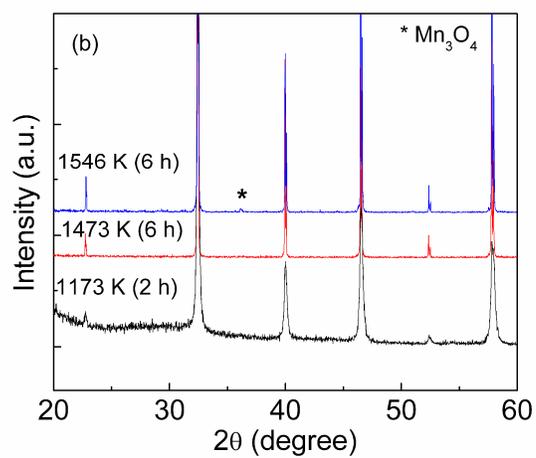

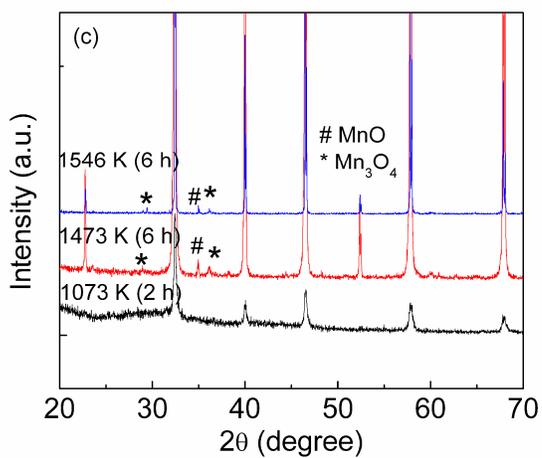

**Fig. 8** High resolution XRD patterns of bulk ceramic samples of **(a)** SMTO5 **(b)** SMTO7 and **(c)** SMTO9 sintered at various temperatures.

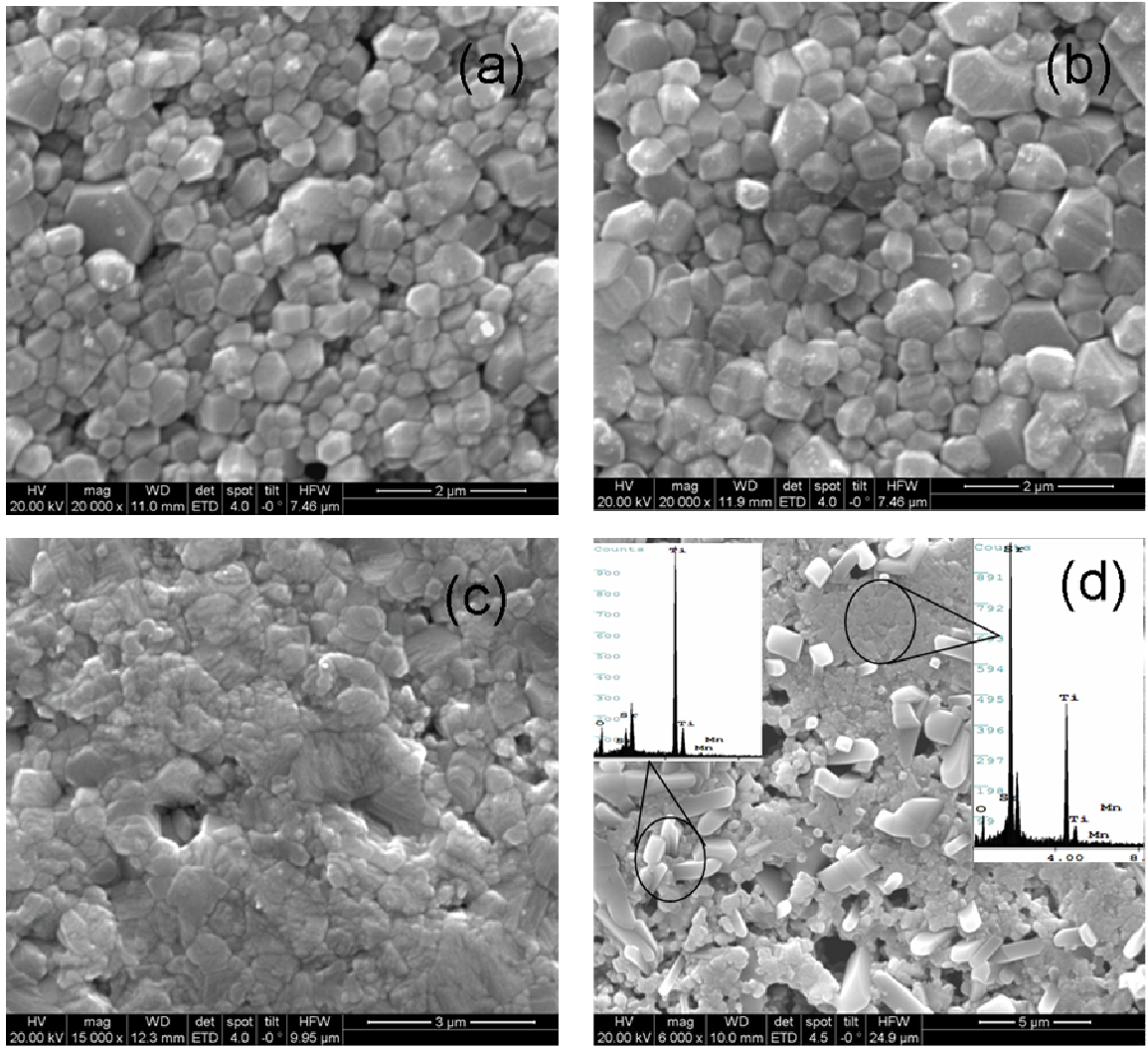

**Fig.9** SEM micrographs of (a) SMTO3 (b) SMTO5 sintered at 1540 K (c) SMTO7 and (d) SMTO9 sintered at 1523 K. The inset shows the EDAX spectra recorded from 2 different spots of the sample.

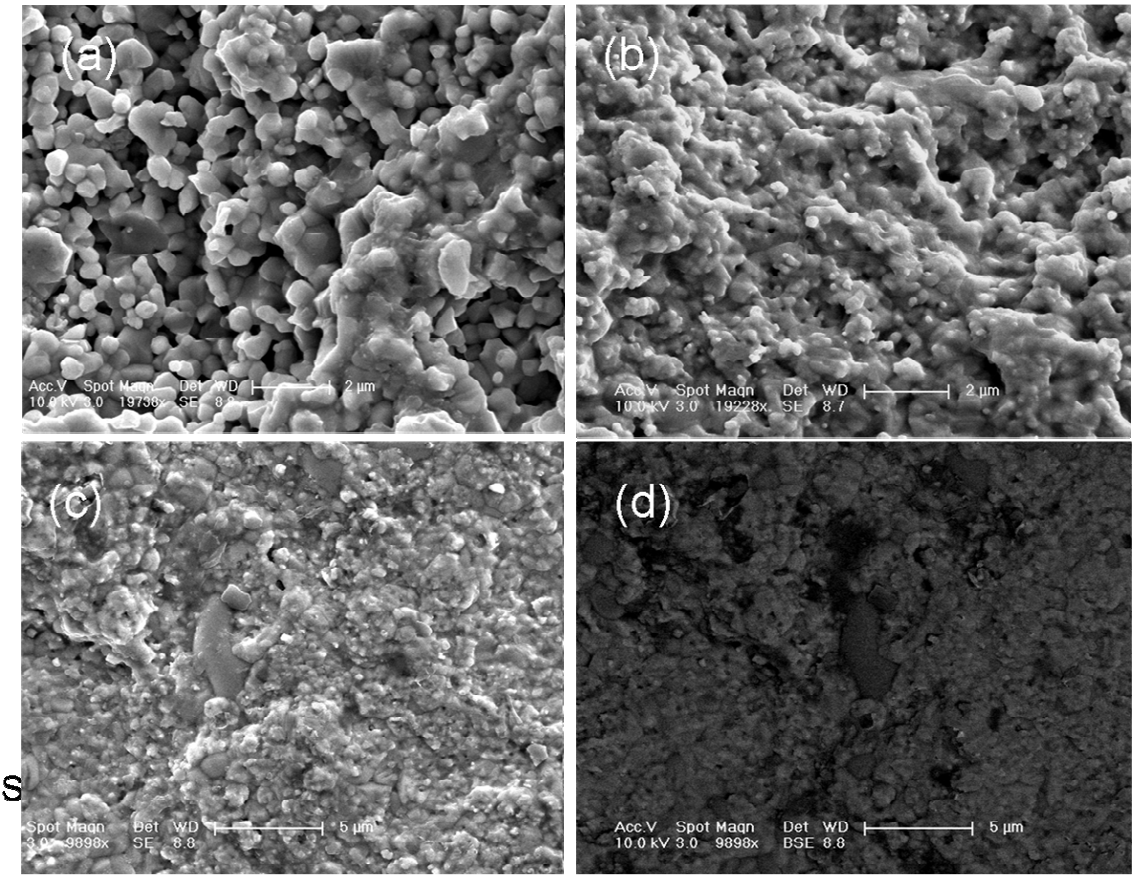

**Fig.10** SEM images of (a) SMTO5 (b) SMTO7 (c) SMTO9 sintered at 1473 K and (d) the corresponding back scattering image of SMTO9 which indicates the segregation of secondary phases.

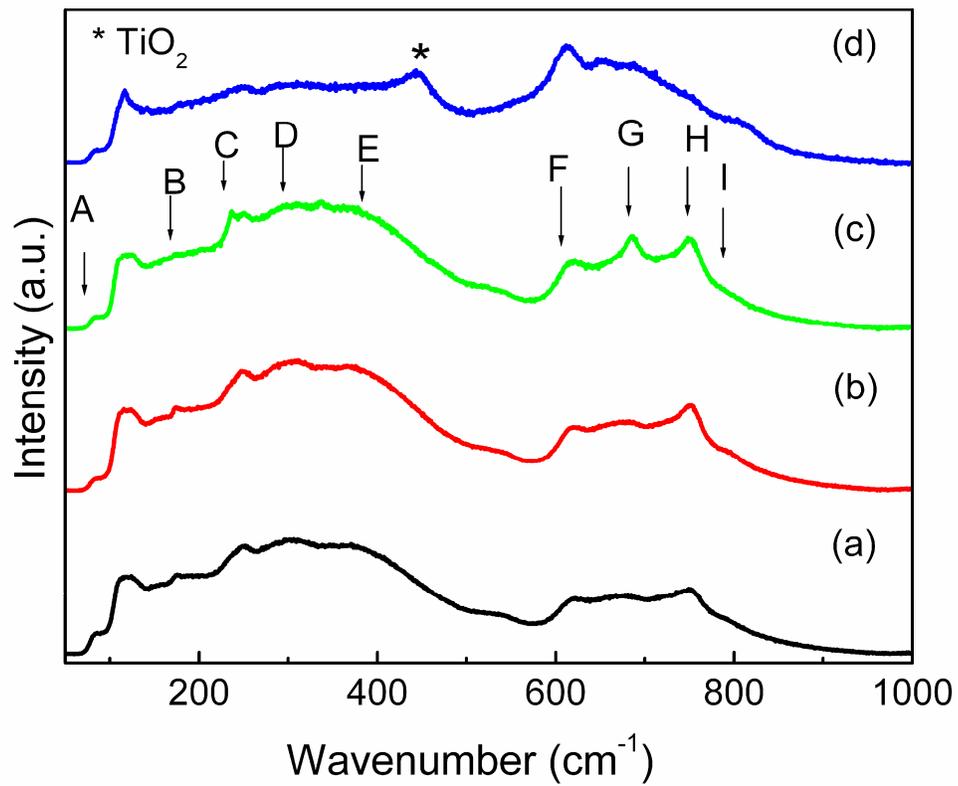

**Fig. 11** Room temperature Micro-Raman spectra of (a) SMTO3 (b) SMTO5 (c) SMTO7 and (d) SMTO9 synthesized using oxalate precipitation technique.

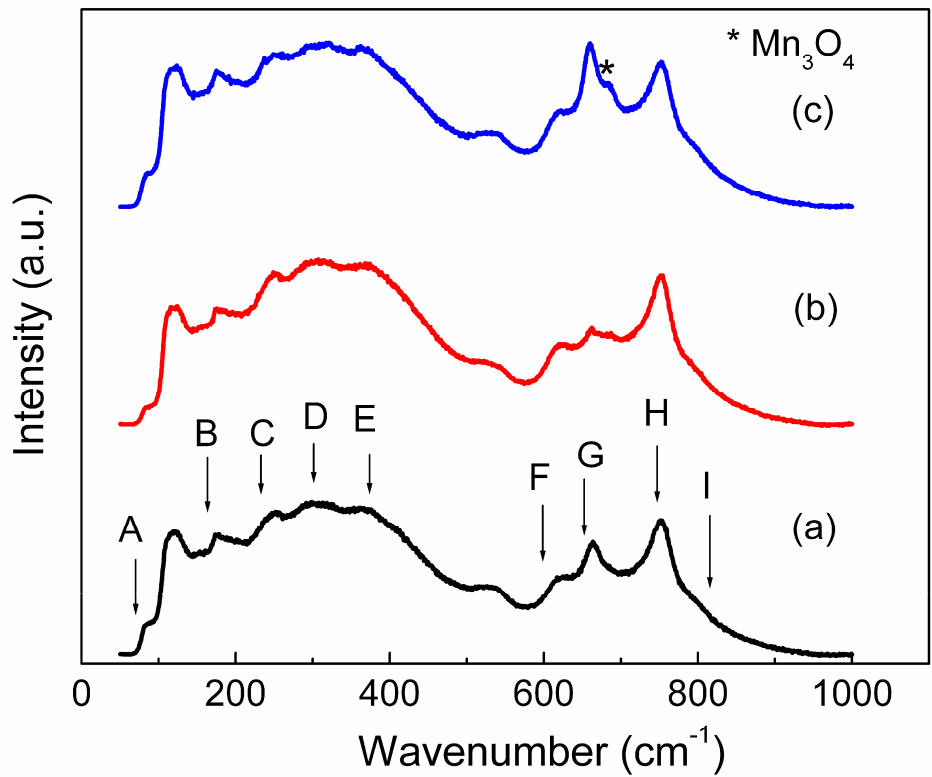

**Fig. 12** Room temperature Micro-Raman spectra of (a) SMTO5 (b) SMTO7 and (c) SMTO9 synthesized using freeze drying method.

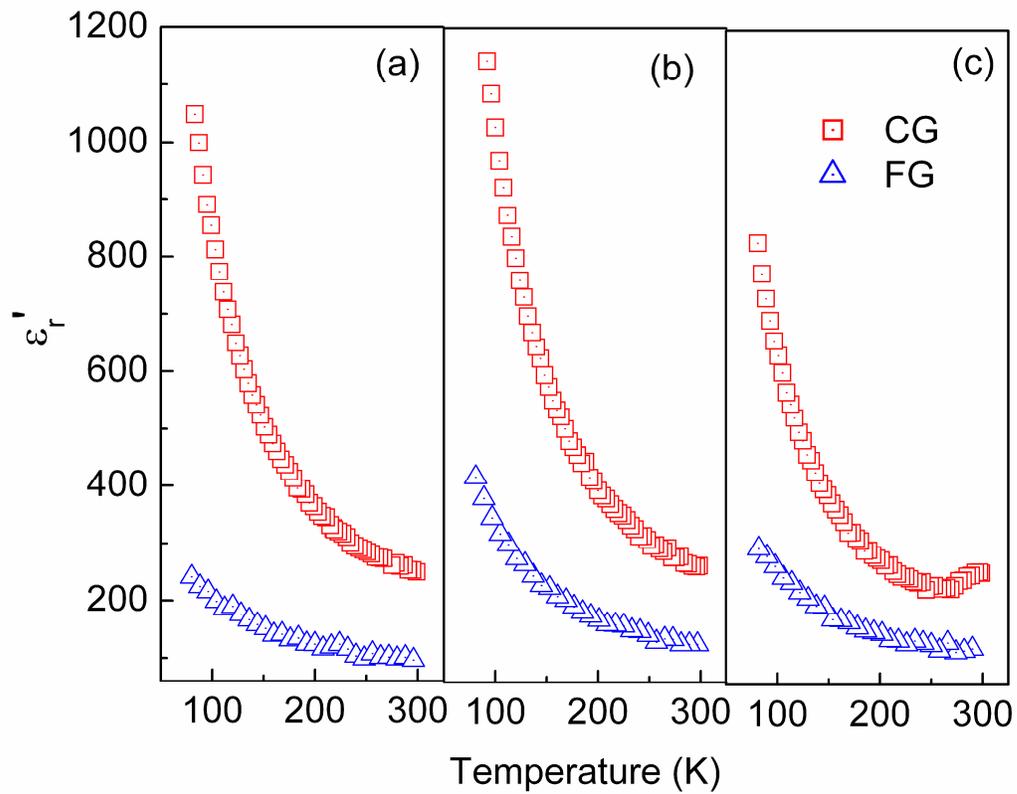

**Fig. 13** Variation of dielectric constant with temperature (80 K to 300 K) at 100 KHz for (a) SMTO5 (b) SMTO7 (c) SMTO9 where open squares represents the ceramics fabricated from oxalate precipitation method (CG) and the open triangles represents the ceramics prepared by freeze drying (FG).